\documentclass[journal]{IEEEtran} 				
\IEEEoverridecommandlockouts 						
\usepackage{amsmath,amssymb}
\usepackage{amsthm}
\usepackage{color}
\usepackage[mathscr]{eucal}
\usepackage{graphics,graphicx,multicol}
\usepackage{epsfig}
\usepackage{enumerate}
\usepackage{subfigure}
\usepackage{algorithmic}
\usepackage{algorithm}
\usepackage{morefloats}
\usepackage{enumerate}
\usepackage{subfigure}
\usepackage{multirow}
\usepackage{array}
\usepackage{pstricks, pst-node, pst-plot, pst-circ}
\usepackage{moredefs}
\usepackage{flexisym}

\usepackage{courier}

\begin{document}

\title{A QoS-based Power Allocation for Cellular Users with Different Modulations}
 \author{\IEEEauthorblockN{Ying Wang\IEEEauthorrefmark{1}, Ahmed Abdelhadi\IEEEauthorrefmark{2}}\\
     \IEEEauthorblockA{\IEEEauthorrefmark{1}EECS, University of Michigan, wying@umich.edu},\\
     \IEEEauthorblockA{\IEEEauthorrefmark{2}Hume Center, Virginia Tech, aabdelhadi@vt.edu}
 }
\maketitle

\begin{abstract}
In this paper, we propose a novel optimal power allocation method that features a power limit function and is able to ensure more users  reach the desired Quality-of-Service (QoS). In our model we use sigmoidal-like utility functions to represent the probability of successful reception of packets at user equipment (UE)s. Given that each UE has a different channel quality and different location from base station (BS), it has different CQI and modulation. For each CQI zone, we evaluate the power threshold which is required to achieve the minimum QoS for each UE and show that the higher CQI the lower power threshold is. We present a resource allocation algorithm that gives limited resources to UEs who have already reached their pre-specified minimum QoS, and provides more possible resources to UEs who can not reach it. We also compare this algorithm with the optimal power allocation algorithm in \cite{abdelhadi2014optimal} to show the enhancement.
\end{abstract}

\pagenumbering{gobble}

\section{Introduction}\label{sec:intro}

In recent years, the user demand for higher data rates and QoS is increasing significantly. The main requirements for the new access network are higher spectral efficiency and higher peak data rates \cite{3gpp}. These needs lead to the existing of 3GPP long term evolution (LTE), the access part of the Evolved Packet System (EPS), to provide higher modulation schemes such as QPSK, 16-QAM, and 64-QAM and. LTE equips the Medium Access Control (MAC) protocol layer as the distributed solution for scheduling \cite{3gpp}. A lot of research work has been done to provide an optimal resource allocation solution for users to seek better QoS. The goal is to provide better signal-to-noise ratio (SNR) and guarantee minimum successful transmission probability of packets. 

The network resource allocation problem can be considered as a maximization of utility functions. The utility function is a representation of each UE's QoS and it is a function of its power allocation. Earlier in \cite{DBLP:journals/corr/Abdel-HadiC14a} and \cite{DBLP:journals/corr/ShajaiahAC14}, the utility function has been approximated as a sigmoidal-like function. The goal is therefore to maximize the network utility which is a product of all users' utilities.

When designing and deploying a wireless network, it is essential to consider the signal coverage. There exists various environments between the BS and UEs, therefore, it is hard to have a unique model to describe the propagation. The path-loss model is the core of signal coverage for any environments \cite{pathloss} and it provides information on the maximum resource that a UE can receive at a distance from the BS. 

During each Transmission Time Interval (TTI), the BS scheduler prioritizes the QoS requirements among the UEs and allocates resources to the UEs based on the information that is feedback from the UEs. This information is the Channel Quality Indicator (CQI) which indicates the perceived quality and the data rate can be supported by the downlink channel. It is carried out when a Block Error Rate (BLER) is smaller than 10\% and the thresholds are set to the SINR values with the BLER smaller than 10\%. Each UE has a different channel quality, i.e CQI value, based on its location from the BS and the environment surrounding it. It was shown in \cite{Wang1602:Optimal} that the sigmoidal-like utility function is a good approximation for the CQI verses power allocated. Therefore, in our paper we represent each CQI with a sigmoidal-like function.

Opportunistic resource allocation algorithms has been proposed in \cite{viswanath2002opportunistic} to improve the system efficiency, however the QoS requirements of users and fairness in allocation failed to be addressed. In our work, we focus on the enhancement of the  resource allocation problem with a sigmoidal-like utility function for each UE. The optimization problem is to achieve the fairness in resource allocation and ensure each UE receives the maximal possible resources to achieve its minimal QoS.

\subsection{Related Work}\label{sec:related}
Early in \cite{harks2005utility}, the authors characterized the resource allocation problem as a global optimization problem and proposed utility proportional fairness criterion to solve this problem. They also showed the bandwidth utility values were ensured to be proportional fair in equilibrium. In \cite{song2005utility}, both utility-based resource management and QoS framework and resource allocation algorithms were studied. The authors also showed an efficient resource allocation for heterogeneous traffic with various QoS requirements. 

The study in \cite{tychogiorgos2011towards} proposed a non-convex optimization algorithm to maximize the utility functions in wireless networks. This optimization framework included a distributed-gradient-based algorithm that solves the optimization problems when the duality gap is zero. In cases of non-zero duality gap, they presented the fair-allocation heuristic and led to an approximated optimal solution. 

In \cite{abdelhadi2014optimal}, the study modeled the user's utility function using sigmoidal like functions and it provided an algorithm for optimal power allocation in a cellular network. Utility proportional fairness was considered and the optimization problem was stated as a product of utilities of all users. In \cite{Haya1} and \cite{Haya2}, a similar approach for optimal power allocation was introduced.

In LTE, the frequency domain scheduler allocates a certain resource block (RB) at a certain transmit rate to a UE basing on the CQI feedback from UEs \cite{donthi2011joint}. In \cite{donthi2011joint}, the authors proposed a trackable model for the CQI feedback schemes in LTE. And in \cite{xie2012dynamic}, the author proposed a dynamic resource allocation algorithm with imperfect channel sensing. Their algorithm keeps tracking the change in channel quality and uses discrete stochastic optimization method to the joint power and channel allocation problem. 

In \cite{hasna2004optimal}, the authors formulated and solved the power allocation problem for multihop transmission. They suggested that the system enhancement is required especially when the highly unbalanced communication links or a large number of hops in the systems, and this enhancement is done by the power optimization. The study in \cite{li2001capacity} proposed an optimal resource allocation for a set of time-invariant additive white Gaussian noise broadcast channels for code division and time division. 

The research work in \cite{ergen2003qos} presented a suboptimal solution that fairly allocates resources and meets the QoS constraints. They showed that the algorithm efficiently converges close to the optimal, and performances well in terms of fair scheduling among users. In \cite{shen2005adaptive}, the study presented a resource allocation framework in multiuser orthogonal frequency division multiplexing (MU-OFDM) systems to achieve variable proportional fairness constraints. This algorithm maximizes the sum channel capacities while maintaining proportional fairness among all UEs.

\subsection{Our Contributions}
The main contributions of this paper are:
\begin{itemize}
\item we proposed a novel optimal power allocation algorithm that includes the power limit feature.
\item we simulated and showed that the optimal power allocation with power limit would ensure more UEs reach the desired QoS, and more power would be allocated to UEs who can not reach the power limit.
\item we compared this algorithm with the optimal power allocation algorithm in \cite{abdelhadi2014optimal} to measure the improvements. 
\end{itemize}

This paper is structured as follow. Section \ref{sec:SM} gives the overview of the system model. In Section \ref{CQI Mapping to Utilities}, we present the process of how we mapped the CQIs to utility functions, and list the resulted parameters corresponding to each CQI along with discussions. The optimal power allocation with power limit algorithm is described in detail in Section \ref{sec:opa}. Section \ref{sec:sim} discusses the simulation results and compares the results with the one of algorithm in \cite{abdelhadi2014optimal}. Finally, Section \ref{sec:conclude} concludes the paper.

\begin{figure}[!t]
\centering
\includegraphics[width=3.5in]{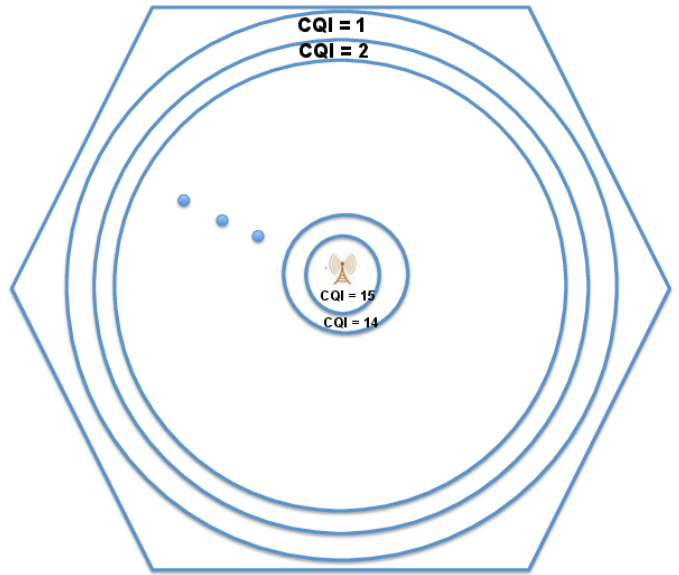}
\caption{System Model}
\label{fig_simu}
\end{figure}

\section{System Model}\label{sec:SM}
In this paper, we consider a single cellular system consisting of a single BS and $M$ UEs. Each UE is placed in a different CQI zone and has a different CQI and corresponding modulation. The set up is shown in Figure \ref{fig_simu}. The UE feedsbacks CQI to indicate the downlink channel quality. It scales from 0 to 15 as shown in Table \ref{tab:title1}. A larger CQI indicates a better channel quality. Based on the CQI information, the BS selects an appropriate modulation scheme and code rate for downlink transmission. The BS distributes its total power $P_{T}$ to all UEs in the cell.

\section{CQI Mapping to Utilities}\label{CQI Mapping to Utilities}
\begin{figure}[!t]
\centering
\includegraphics[width=3.5in]{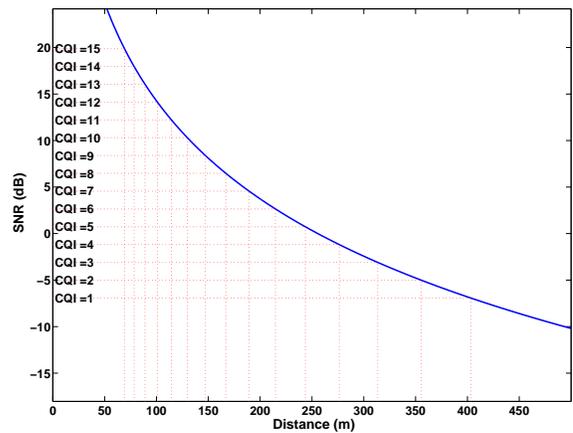}
\caption{Mapping CQI, SNR with distance from the BS}
\label{fig_CQI_DIST}
\end{figure}

The  path-loss is calculated in (\ref{eqn:Pathloss}) to map each CQI to its corresponding distance from BS as shown in Figure \ref{fig_CQI_DIST}. $\alpha$ in (\ref{eqn:Pathloss}) is the the path loss exponent and in a urban environment it equals to 3.5. Therefore each CQI zone will have different distances from the BS as well as the UEs' location. 

\begin{equation}\label{eqn:Pathloss}
P_{\text{UE}} = \frac{P_{\text{BS}} f}{c (4 \pi d)^\alpha} 
\end{equation}\\
where $f$ is the carrier frequency and $c$ is speed of the light.\\

\begin{figure}[!t]
\centering
\includegraphics[width=3.5in]{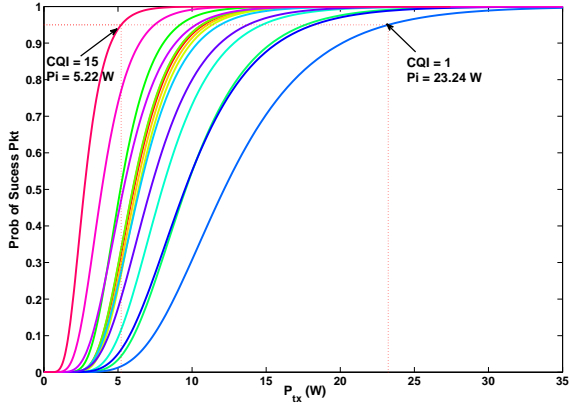}
\caption{Power Utility Function}
\label{fig_power_utility}
\end{figure}

The probability of the successful package reception is calculated from the efficiencies of different CQI values in Table \ref {tab:title1}. The utility function as a result of CQI mapping is shown in Figure \ref{fig_power_utility}. It is a function of the transmitter power. It looks like sigmoidal-like function, therefore we can use the normalized sigmoidal-like utility function, as in \cite{DBLP:journals/corr/Abdel-HadiC14a} and \cite{DBLP:journals/corr/ShajaiahAC14}:

\begin{equation}\label{eqn:sigmoid}
U_i(P_i) = c_i\Big(\frac{1}{1+e^{-a_i(P_i-b_i)}}-d_i\Big)
\end{equation}
where $c_i = \frac{1+e^{a_ib_i}}{e^{a_ib_i}}$ and $d_i = \frac{1}{1+e^{a_ib_i}}$. \\

To curve fit we used the Levenberg-Marquardt (LM) Optimization method to identify the parameters in the utility functions for different CQIs. The evaluated parameters of 15 utility functions are listed in Table \ref{tab:title1}. The left four columns in Table \ref{tab:title1} are the standard LTE CQI coefficients. The values for $a$s and $b$s in column 5 and 6 are the parameters in (\ref{eqn:sigmoid}).

For example, in Figure \ref{fig_power_utility}, we set the minimum QoS requirement to achieve at least a 95\% successful packet transmission. For UE with CQI 15, the power that required to achieve the minimum QoS is about 5.22W whereas the UE with CQI 1 needs 23.24W to have the minimum QoS. A lower amount of power is required to achieve the desired QoS by the UE with a better channel quality, and this fact motivates us to include a power limit in our optimal power allocation algorithm.

\begin{table}[!t]
\renewcommand{\arraystretch}{1.3}
\caption{Utility parameters} \label{tab:title1} 
\begin{tabular}{ | l | l | l | l | l | l | l |}
\hline
\shortstack{ \\CQI \\Index} & Modulation & \shortstack{ \\Code Rate \\ X 1024} & Efficiency & a & b \\\hline
1 & QPSK  & 78 & 0.1523 & 0.8676 & 6.2257 \\ \hline
2 & QPSK & 120 & 0.2344 & 0.8761 & 6.1657 \\ \hline
3 & QPSK & 193 & 0.3880 & 0.8466 & 6.3812 \\ \hline
4 & QPSK & 308 & 0.6016 & 0.8244 & 6.5526 \\ \hline
5 & QPSK & 449 & 0.8770 & 0.8789 & 6.1467 \\ \hline
6 & QPSK & 602 & 1.1758 & 1.0188 & 5.3029 \\ \hline
7 & 16QAM & 378 & 1.4766 & 0.5077 & 9.8303 \\ \hline
8 & 16QAM & 490 & 1.9141 & 0.6086 & 8.1999 \\ \hline
9 & 16QAM & 616 & 2.4063 & 0.7524 & 6.6333 \\ \hline
10 & 64QAM & 466 & 2.7305 & 0.3697 & 12.5005\\ \hline
11 & 64QAM & 567 & 3.3223  & 0.4722 & 9.7873\\ \hline
12 & 64QAM & 666 & 3.9023 & 0.6248 & 7.3974 \\ \hline
13 & 64QAM & 722 & 4.5234& 0.8376 & 5.5177\\ \hline
14 & 64QAM & 873 & 5.1152  & 1.1510 & 4.0153\\ \hline
15 & 64QAM & 948 & 5.5547  & 1.6471 & 2.8058\\
\hline
\end {tabular}
\end{table}

\section{Optimal Power Allocation With Power Limit}\label{sec:opa}
In Section \ref{sec:SM}, we discussed that the user with a better channel quality will require less power to achieve the same QoS than the user with a worse channel quality. In this section, we introduce a robust distributed algorithm with power limits.

\subsection{Problem Formulation}\label{sec:PF}
We form the utility proportional fairness power allocation problem as following:
\begin{equation}\label{eqn:opt}
\begin{aligned}
& \underset{\textbf{P}}{\text{max}}
& & \prod_{i=1}^{M}\log(U_i(\gamma_i(P_i))) \\
& \text{subject to}
& & \sum_{i=1}^{M}P_i \leq P_T\\
& & &  P_i \geq 0, \;\;\;\;\; \text{for} \; \; i = 1,2, ...,M \; \; \text{and} \; \; P_T \geq 0.
\end{aligned}
\end{equation}
where $P_T$ is the total power of the BS, $M$ is the number of UEs and $\textbf{P} = \{P_1, P_2, ..., P_M\}$.\\

Given that the objective function in (3) is strictly concave, the optimization problem is convex \cite{abdelhadi2014optimal} and therefore there exists a unique tractable global optimal solution.

\subsection{Robust Distributed Algorithm with Power Limits}
The power limit is the transmitter power that a UE requires to achieve his/her QoS. In our model, we assume the minimum QoS is to reach 95\% successful package, and the pre-specified power limits are the amount of power required to achieve this QoS. The algorithm is shown in Algorithm (\ref{alg:UE_FK}) and (\ref{alg:eNodeB_FK}):
\begin{algorithm}
\caption{UE Algorithm}\label{alg:UE_FK}
\begin{algorithmic}
\STATE {Send initial bid $w_i(1)$ to eNodeB}
\LOOP
  \STATE {Receive shadow price $p(n)$ from eNodeB}
  \IF {STOP from eNodeB} %
    \STATE {Calculate allocated rate $P_i ^{\text{opt}}=\frac{w_i(n)}{p(n)}$}
    \STATE {STOP}
  \ELSE
    \STATE {Solve $P_{i}(n) = \arg \underset{{P_{i}}}{\text{max}}(\log U_{i}(\gamma_{i}(P_{i}))-p(n)P_{i})$}
    \IF {$P_{i} > PowerLimit_{i}$}%
          \STATE {Allocate rates: $P_{i}^{\text{opt}} = P_{i}$ and $P_{T} = P_{T} - P_{i}$ to user $i$}
          \STATE {$UE_{i}$ quits the Optimal Power allocation and rest UEs continue to bid}
    \ENDIF 
     \STATE {Calculate new bid $w_i(n) = p(n)P_{i}(n)$ for remaining UEs}
     \IF {$|w_i(n)-w_i(n-1)| >  \Delta w(n)$}  %
      \STATE{$w_i(n) = w_i(n-1)+\text{sign}(w_i(n)-w_i (n-1)) \Delta w(n)$}
      \STATE{$\big\{\Delta w(n) = l_1 e^{n/l_2} \big \}$}
     \ENDIF 
     \STATE {Send new bid $w_i (n)$ to eNodeB}
  \ENDIF 
\ENDLOOP
\end{algorithmic}
\end{algorithm}
\begin{algorithm}
\caption{eNodeB Algorithm}\label{alg:eNodeB_FK}
\begin{algorithmic}
\LOOP
  \STATE {Receive bids $w_i(n)$ from UEs}
  \COMMENT{Let $w_i(0) = 0\:\:\forall i$}
      \IF {$|w_i(n) -w_i(n-1)|< \delta  \:\:\forall i$} %
        \STATE {Allocate rates:$P_{i}^{\text{opt}}=\frac{w_i(n)}{p(n)}$ to user $i$} 
        \STATE {STOP} 
    \ELSE
       \STATE {Calculate $p(n) = \frac{\sum_{i=1}^{M}w_i(n)}{P_{T}}$} 
       \STATE {Send new shadow price $p(n)$ to remaining UEs}
  \ENDIF 
\ENDLOOP
\end{algorithmic}
\end{algorithm}

The algorithm is divided into a UE algorithm shown in Algorithm (\ref{alg:UE_FK}) and a BS algorithm shown in Algorithm (\ref{alg:eNodeB_FK}). Each UE starts sending an initial bid $w_{i}(1)$ to the BS. The BS calculates the difference between the received bid $w_{i}(n)$ and the previously received bid $w_{i}(n-1)$ and compares its value to a pre-specified threshold $\delta$. If it is greater than the threshold $\delta$, the BS calculates the shadow price $p(n) = \frac{\sum_{i=1}^{M}w_i(n)}{R}$ and sends it to UEs. Each UE receives the shadow price $p(n)$ from the BS and solves the power $P_{i}$ that maximize $(\log U_{i}(\gamma_{i}(P_{i}))-p(n)P_{i})$, then compares $P_{i}$ to its power limit $PowerLimit_{i}$ and the user stays in the process if $P_{i} < PowerLimit_{i}$. If it is greater the $UE_{i}$ exists the power allocation and $P_{i}$ is subtracted for the total power $P_{T}$. After that each remaining UE calculates a new bid $w_i(n) = p(n)P_{i}(n)$ and decreases the difference between the current bid 
and previous bid $w_{i}(n) - w_{i}(n-1)$ using exponential function $\Delta w(n) = l_1 e^{n/l_2}$. The reason that we use this exponential function is that when ${\sum_{i=1}^{M }P_{i}^{inf}  = \sum_{i=1}^{M }b_{i}} \geq {P_{T} }$ the convergence to the optimal powers can no longer be guaranteed as it fluctuates about the global optimal solution. Therefore the exponential fluctuation decay function is introduced to resolve the problem. Each remaining UE sends the new bid $w_i(n) = w_i(n-1)+ \text{sign}(w_i(n)-w_i (n-1)) \Delta w(n)$ to BS. This process repeats until $|w_i(n) -w_i(n-1)|$ is less than the pre-specified threshold $\delta$.

\section{Simulation Results}\label{sec:sim}
\begin{figure}[!t]
\centering
\includegraphics[width=3.5in]{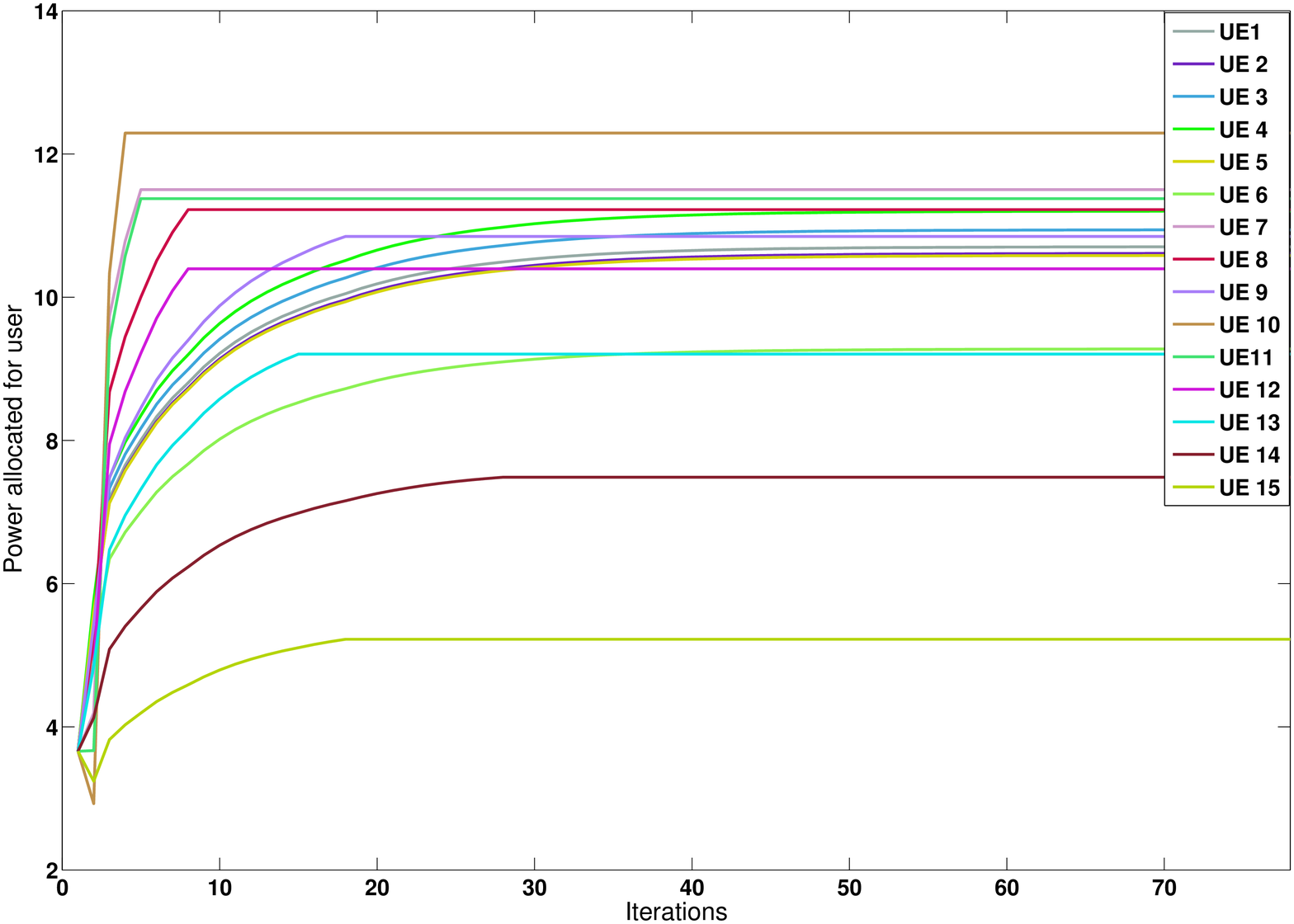}
\caption{Power allocation to 15 different CQI users at BS power 15W using algorithm with PL}
\label{fig_PA_PL_power}
\end{figure}

\begin{figure}[!t]
\centering
\includegraphics[width=3.5in]{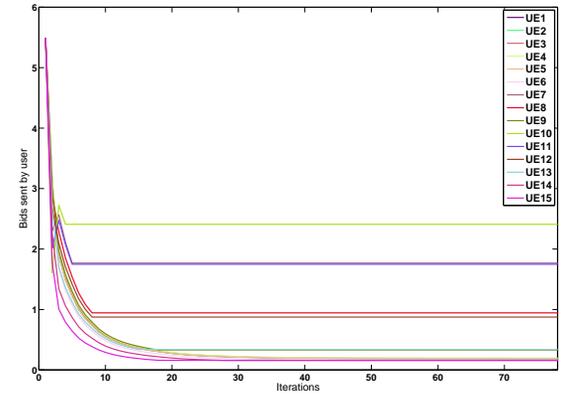}
\caption{Bids sent by 15 different CQI users at BS power 150W using algorithm with PL}
\label{fig_PA_PL_bids}
\end{figure}

The BS has total power $P_{T}$ = 150W to distribute to 15 UEs. Each UE stands in a different CQI zone and is represented by a sigmoidal-like utility function. Algorithm (\ref{alg:UE_FK}) and (\ref{alg:eNodeB_FK}) were simulated in MATLAB. The simulation results showed that the optimal powers were allocated to all users as shown in Figure \ref{fig_PA_PL_power}. The bidding process for 15 UEs is plotted in Figure \ref{fig_PA_PL_bids}. 

\begin{table}[!t]
\renewcommand{\arraystretch}{1.3}
\caption{Comparison between Power Allocation with and without Power Limits} \label{tab:PL} 
\begin{tabular}{ | l | l | l | l |l|}
\hline
UE &\shortstack{ \\ Power to \\reach QoS (W)} & \shortstack{Power with\\ PL (W)} & \shortstack{\\Power without\\ PL (W)}& \shortstack{ \\Reach desired\\ QoS}\\ \hline
1 & 23.240 & 10.491 & 9.122 & No\\ \hline
2 & 18.210 & 10.401 & 9.045 & No\\ \hline
3 & 17.650 & 10.723 & 9.318 & No\\ \hline
4 & 14.720 & 10.978 & 9.5337 & No\\ \hline
5 & 13.760 & 10.373 & 9.0218 & No\\ \hline
6 & 11.910 & 9.0968 & 7.9388 & No\\\hline
7 & 11.350 & 11.502 & 13.6145 & Yes\\ \hline
8 & 11.060 & 11.223 & 11.6935 & Yes\\ \hline
9 & 10.790 & 10.849 & 9.7709 & Yes\\ \hline
10 & 10.690 & 12.291 & 16.7008 & Yes\\ \hline
11 & 10.650 & 11.376 & 13.6879 & Yes\\ \hline
12 & 10.260 & 10.397 & 10.8536 & Yes\\ \hline
13 & 9.181 & 9.2056 & 8.4798 & Yes\\ \hline
14 & 7.485 & 7.4862 & 6.4664 & Yes\\ \hline
15 & 5.213 & 5.2229 & 4.7468 & Yes\\ 
\hline
\end {tabular}
\end{table}

We also provided a comparison of the allocated power for each UE between the optimal power allocation algorithm with power limit and without it \cite{abdelhadi2014optimal} in Table \ref{tab:PL}. We assumed the minimal QoS is to reach at least a 95\% success package transmission at UEs. Column 2 in Table \ref{tab:PL} indicates the power that required by each UE to achieve the minimal QoS, and we set those values to be the power thresholds. For example, UE 15 requires 5.213W to reach the minimal QoS, after it receives 5.2229W from the BS it quits the algorithm and the remaining power is allocated to the rest of the UEs. There are more UEs achieving their minimal QoS, e.g. UE 13 receives 9.2056W and achieves the desired QoS using our algorithm while it only receives 8.4798W and fails to meet the QoS with the algorithm without power limit.  With our new algorithm there are 9 UEs achieving the minimal QoS while the algorithm without power limit in \cite{abdelhadi2014optimal} only has 5 UEs meeting the QoS 
requirements. Even for those UEs who do not reach the desired QoS, our algorithm still allocates higher power to them than the algorithm without power limit. For example, UE 1 receives 10.491W by using our algorithm while 9.122W is allocated to it with the algorithm without the power limit.

\section{Conclusion}\label{sec:conclude}
In this paper, we proposed a new optimal power allocation algorithm with power limit feature that ensures more UEs achieve the desired QoS and guarantees more power to be allocated to UEs that can not meet the QoS requirements. We simulated this algorithm with one BS and 15 different CQI UEs. We showed that this new algorithm allows more users to reach their desired QoS, and at the same time more power is allocated to the users who do not reach the power limits comparing to the algorithm in \cite{abdelhadi2014optimal}.

\bibliographystyle{ieeetr}
\bibliography{paper2cit}
\end{document}